# High-resolution multimodal flexible coherent Raman endoscope

**Alberto Lombardini[1], Vasyl Mytskaniuk[1], Siddharth Sivankutty[1], Esben Ravn Andresen[1,2], Xueqin Chen[1], Jérôme Wenger[1], Marc Fabert[4], Nicolas Joly[3], Frédéric Louradour[4], Alexandre Kudlinski[2], and Hervé Rigneault[1*]**

[1] Aix Marseille Univ, CNRS, Centrale Marseille, Institut Fresnel, Marseille, France

[2] Université Lille, CNRS, UMR 8523, Laboratoire de Physique des Lasers Atomes et Molécules, F-59000 Lille, France

[3] Max Planck Institute for the Science of Light, and Department of Physics, University of Erlangen Nuremberg, 91058 Erlangen, Germany

[4] Université de Limoges, CNRS, XLIM, UMR 7252, Université de Limoges, F-87060 Limoges, France

[*]herve.rigneault@fresnel.fr

**Coherent Raman scattering microscopy is a fast, label-free and chemically specific imaging technique that has a high potential for future in-vivo optical histology. However, its imaging depth into tissues is limited to the sub-millimeter range by absorption and scattering. Performing coherent Raman imaging in a fiber endoscope system is a crucial step to image deep inside living tissues and provide the information inaccessible with current microscopy tools. However the development of coherent Raman endoscopy has been hampered by several issues in the fiber delivery of the excitation pulses and signal collection. Here, we present a flexible, compact, and multimodal nonlinear endoscope (4.2 mm outer diameter, 71 mm rigid length) based on a resonantly scanned hollow-core Kagomé-lattice double-clad fiber. The fiber design allows distortion-less, background-free delivery of femtosecond excitation pulses and the back-collection of nonlinear signals through the same fiber. Sub-micron spatial resolution together with large field of view is made possible by the combination of a miniature objective lens together with a silica microsphere lens inserted into the fiber core. We demonstrate coherent anti-Stokes Raman scattering, 2-photon fluorescence and second harmonic generation imaging with 0.8 μm resolution over a field of view up to 320 μm and at a rate of 0.8 frames/s. These results pave the way for intra-operative label-free imaging applied to real-time histopathology diagnosis and surgery guidance.**

The identification of cancer tumors is generally performed *ex vivo* by the histological inspection of tissue biopsies. This process requires several steps such as tissue resection, sectioning and staining, and is hence time-consuming and labour-intensive. Real-time label-free in-vivo endoscopy imaging would realize a major breakthrough in histology. By enabling direct visualization of cancerous cells and tissues inside the patient's body, such an endoscope system would revolutionize the current approaches for cancer diagnosis and intraoperative surgery decision-making.

As a significant step towards this goal, coherent anti-Stokes Raman scattering (CARS) and stimulated Raman scattering (SRS) are fast, label-free and chemically specific imaging techniques[1, 2] that bear a high potential for in-vivo intraoperative histopathology[3, 4, 5]. However, despite the capacity of CARS and SRS to provide high-resolution microscopy images of unprocessed samples, these techniques cannot image deeper than a few hundreds of micrometers inside tissues as they are limited by optical absorption and scattering[6]. Moreover, CARS and SRS are implemented on unwieldy microscope setups on optical tables, and so miniaturization and portability remain challenging issues for clinical applications.

Performing CARS and/or SRS imaging in a fiber endoscope system is a crucial step to image deep inside tissues and provide information that is inaccessible with current microscopy tools. Despite significant efforts over the last decade[7, 8, 9, 10, 11, 12, 13], major technical challenges in the pulsed laser delivery and signal collection have hindered the development of coherent Raman endoscopy. CARS/SRS endoscopy is more demanding than 2-photon or second harmonic generation (SHG)[14, 15] as two spatially and temporally overlapping excitation beams are required. When these two pulses co-propagate in the fiber, a strong parasitic background arises due to nonlinear four wave mixing (FWM) occurring in the fiber core[7]. Moreover, group velocity dispersion (GVD) and a variety of nonlinear effects affect the pulse propagation in fibers, leading to pulse temporal and spectral broadening and a loss of peak intensity[16]. For CARS and SRS, GVD and pulse propagation issues have to be compensated at two different wavelengths, which further complicates the challenge. To deal with these problems, earlier works used different fibers for excitation and collection[7, 11, 13], or dedicated fibers with large mode area[9] or polarization control[8]. However, the complexity introduced by these designs leads to bulky endoscope heads and/or a loss in optical resolution and contrast, which are not compatible with in-vivo histopathology applications.

In this letter, we present a compact flexible fiber-optic scanning endoscope dedicated to high-resolution coherent Raman imaging deep inside tissues. Our specific design combines several key innovations: (1) a broadband hollow-core (HC) fiber with a Kagomé lattice guides the two co-propagating laser beams with negligible FWM background, GVD and pulse distortion. (2) A double cladding around the HC fiber with a high numerical aperture provides efficient signal back-collection even with scattering samples. (3) A silica microsphere inserted into the output facet of the HC fiber core solves the issue related to the large mode surface of the HC fiber, and enables a sub-micron spatial resolution for CARS imaging. (4) A miniature resonant piezo scanner combined with a distal miniature objective provides an electrically-tunable field of view (FoV) of a few hundreds of μm while preserving the endoscope distal head compactness (4.2 mm outer diameter, 71 mm rigid length). This unique combination of features enables high-contrast and high-resolution multimodal nonlinear endoscopy of unstained tissues. We perform CARS, SHG and 2-photon endoscopy imaging with 0.8 μm resolution over a FoV up to 320 μm and at a rate of 0.8 frame/s. By solving most of the issues raised in nonlinear endoscopy, our dedicated design opens a major route towards the clinical application of deep in-vivo CARS imaging for real-time histopathology diagnosis.

**Endoscope system overview**

Our multimodal nonlinear endoscope setup is shown in **Fig. 1**. A tunable multi-wavelength laser system (Discovery, Coherent) provides two synchronized fs ultrashort 80 MHz pulse trains for the pump (800 nm – 100 fs) and Stokes (1040 nm – 160fs) beams. These wavelengths are suitable to address carbon hydrogen (C-H) bonds (~2885 $cm^{-1}$) with CARS. The two co-propagating laser pulses are temporally overlapped with a mechanical delay line, and injected into the same fiber core. The laser spot at the fiber distal facet is further re-imaged onto the sample through a custom designed miniature objective based on four commercial achromatic doublets (**Fig. S1**). The magnification of the miniature objective is 0.63, with 0.3 numerical aperture on the fiber side and 0.45 on the sample side. The objective working distance is 0.6 mm in air, and is sufficiently long for tissue imaging applications.

For image scanning, the distal end of the fiber is attached to a four-quartered piezo tube (PT230.94, Physik Instrument – diameter 3.2 mm), with a free standing length of 23 mm. The excitation spot is scanned over the sample by resonant driving of the piezo tube in an expanding spiral pattern (**Fig. S2**). The FoV of the endoscope distal probe is a circle of maximum diameter 320 μm, which results from the de-magnification of the 510 μm fiber scan (at 30 V driving voltage) through the miniature objective lens. The whole endoscope distal head is integrated in a biocompatible stainless steel tube (4.2 mm diameter – length 71.2 mm) which ensures its portability (**Fig. 1 inset picture**).

The optical signal generated in the sample is back-collected through the same objective and fiber. After propagation through the fiber, it is spectrally filtered by a set of dichroic mirrors and bandpass filters and detected by photomultiplier units (H7421-40, Hamamatsu). CARS, SHG and TPEF images are built and displayed in real time with a custom Labview software, based on an open-loop image reconstruction algorithm (**Fig. S3, Fig. S4**).

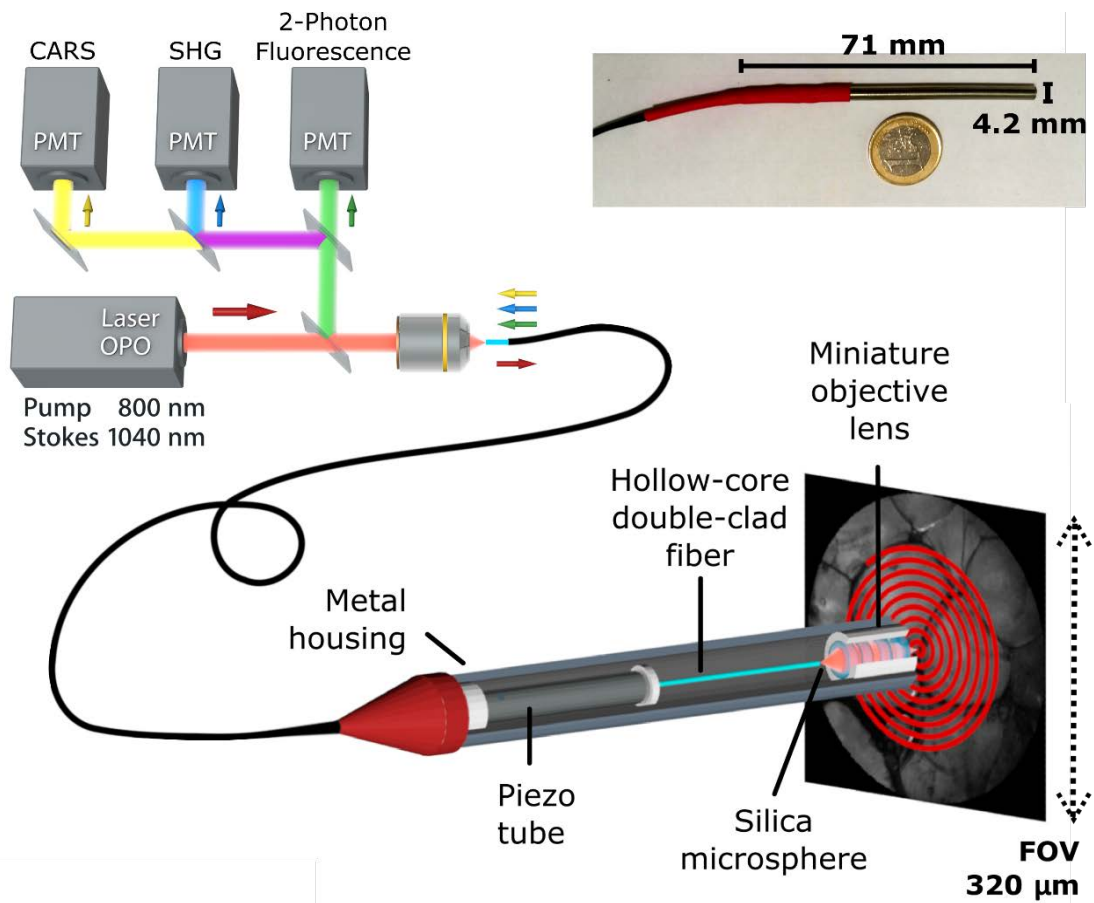

**Figure 1: Coherent Raman endoscope overview.** A tunable femtosecond laser provides two synchronized pulse trains which are injected into the hollow core fiber probe to perform CARS, SHG and 2-photon fluorescence endoscopy imaging. The light emitted by the sample is collected and back-propagates through the same fiber, and is detected on large area photomultiplier detectors. The inset picture shows the endoscope probe head inserted in its stainless steel housing.

**Double-clad hollow-core fiber probe**

One of the key element of our nonlinear multimodal endoscope is the hollow-core fiber[17] **(Fig. 2a, b)** which combines a Kagomé-lattice hollow core (HC) for distorsionless femtosecond pulse delivery with a high numerical aperture all silica double-clad (DC) for nonlinear signal collection. The fiber has an outer diameter of 327 µm and a core diameter of 20 µm where light propagates in a single transverse mode. Light confinement inside the Kagomé core is not the result of a photonic band gap, but is due to a mechanism of inhibited coupling between the cladding modes and the guided core modes[18]. The transmission window extends from 700 nm to 1100 nm with propagation losses below 5 dB/m **(Fig. 2c)**, largely enough to accommodate for both the pump and Stokes beams for CARS. Importantly, the GVD is very weak and remains below 5 fs/nm/m over the full transmission window, hence the temporal broadening of a ~150 fs pulse travelling over the 1 m fiber length is only a few fs and can be fully neglected **(Fig. 2d)**. Additionally, we observe negligible spectral changes after the propagation through the fiber, even for powers up to 60 mW **(Fig. S5)**. Altogether, these results show that the hollow-core fiber is suitable to deliver high-power femtosecond laser pulses with negligible distortion over a broad spectral window.

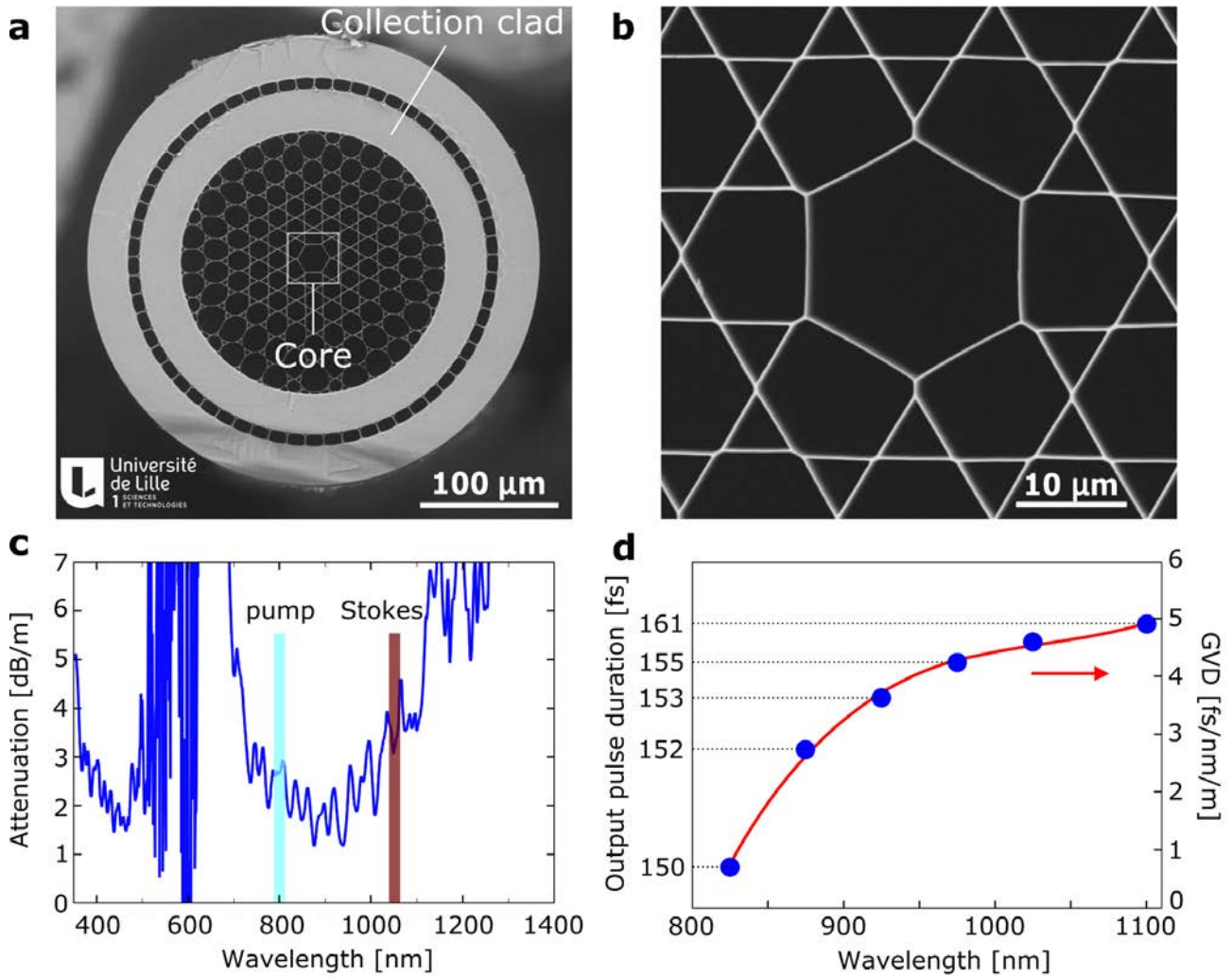

**Figure 2: Hollow core photonic crystal fiber dedicated for multimodal nonlinear endoscopy.** (a) Electron microscope image of the hollow core (HC) fiber featuring a Kagomé lattice together with a double cladding (DC) separated by air holes. The excitation pulses propagate through the HC while the nonlinear signal is collected and transmitted through the DC. (b) Close-up view of the fiber core area. (c) Attenuation of the HC, the wavelengths used for the pump and Stokes beams are highlighted. (d) Output pulse duration after 1 m propagation for a 150 fs input pulse. The output pulse duration are computed using the experimentally values for the GVD (blue dots - right scale), which are deduced from the measurement of the group delay versus wavelength. The red line is a third-order polynomial fit to the GVD data.

The second main challenge for fiber design in coherent Raman imaging is the parasitic background due to nonlinear four wave mixing (FWM) occurring in the fiber core. **Figures S6** and **S7** show that the HC of our fiber design heavily suppresses the parasitic FWM background, enabling noise-free CARS detection. This is in sharp contrast with most solid core fibers that necessitate FWM filtering[11, 13].

The third challenge for CARS endoscopy is the ability to collect the nonlinear signal even in the presence of scattering sample. To solve this issue, our fiber design features a second annular silica cladding dedicated to the collection of the generated nonlinear signals **(Fig. 2a)**. Broadband and multimode guidance in this silica double clad (DC) is obtained by adding a low index ring of air holes (air clad) outside the Kagomé structure. The inner and outer diameters of this cladding are 190 μm and 250 μm respectively, providing an efficient collection surface of ~21000 μm$^2$ with a large ~0.5 numerical aperture. This DC is necessary to collect the CARS, SHG and 2-photon fluorescence whose wavelengths are outside the hollow core transmission window. Its high numerical aperture also increases the endoscope collection efficiency when working with scattering samples such as biological tissues (**Fig. S8**).

**Microsphere lens focusing for hollow-core fiber probe**

So far, we have shown that Kagomé-lattice hollow core fibers have major advantages for nonlinear endoscopy. However, they also feature one significant drawback related to their large guided mode diameter which is inappropriate for high-resolution nonlinear imaging. To overcome this issue, we use a 30 μm silica microsphere inserted into the hollow fiber core (**Fig. 3a, b**). The microsphere acts as a ball lens to strongly focus the ~15 μm diameter guided mode into a ~1 μm focus spot thanks to the so-called photonic jet focusing[19, 20, 21] (**Fig. 3c, d**). This focus spot is then imaged onto the sample plane with the miniature objective which acts as a relay lens (**Fig. 1**). This is the key to achieve submicron image resolution and large FoV as we will show hereafter. To maintain the microsphere in contact with the fiber at high piezo scanning speed, the microsphere was permanently sealed to the fiber by using a $CO_2$ laser splicer. Thanks to the high damage threshold and low absorption of silica, excitation powers higher than 100 mW could be used with no visible impact on the device performance. We found also that the microsphere induces negligible FWM when performing CARS (**Fig. S7**).

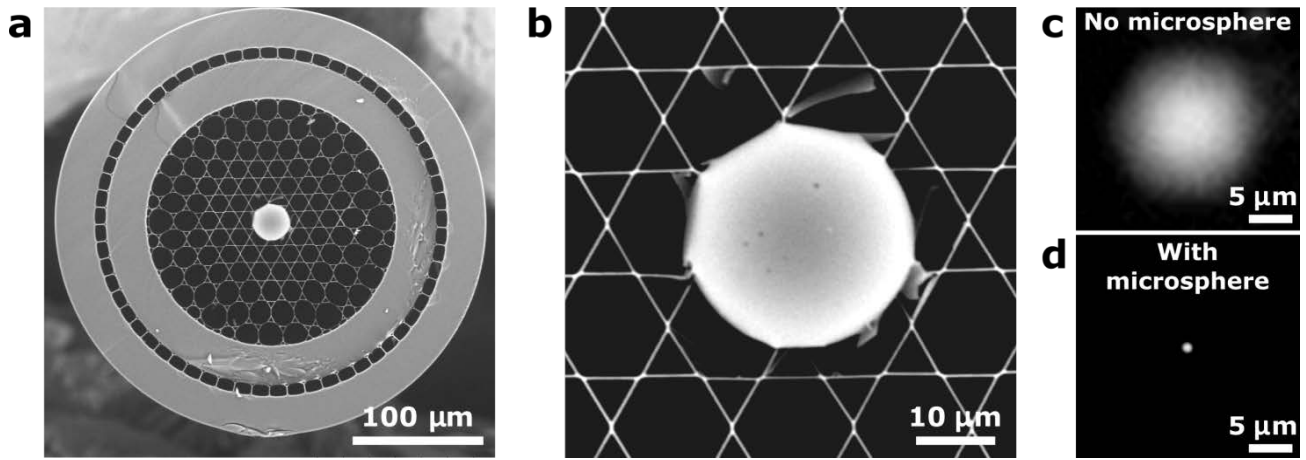

**Figure 3: Microsphere lens inserted into the hollow fiber core provides a submicron focus spot required for imaging**. (a) Scanning electron microscope image of the HC fiber with a 30 µm silica microsphere inserted and sealed into its fiber core. (b) Close-up view of the microsphere inserted into the hollow core. (c) In the absence of the microsphere, the HC mode diameter is 15 µm, which is inappropriate for high-resolution nonlinear imaging. (d) Thanks to the microsphere lens, the light exiting the fiber core is focused into a ~1 µm diameter spot.

**Image quality check**

We first demonstrate the imaging capabilities of our endoscope system on 5 µm polystyrene beads deposited on a glass coverslip. **Figures 4a, b** show CARS images acquired with two different voltages driving the piezo scanner, providing different FoVs of 155 µm and 25 µm respectively. Each bead leads to a strong CARS signal thanks to the 2851 cm$^{-1}$ and 2902 cm$^{-1}$ Raman band of polystyrene, corresponding to aliphatic C-H stretching vibrations. The background between the beads is remarkably dark, which demonstrates CARS imaging free of fiber FWM background.

To calibrate the lateral and axial resolution of our endoscope probe, we record the 2-photon fluorescence image of 200 nm fluorescent nanoparticles excited by a laser pulse at 800 nm (**Fig. 4c**). The image of an individual nanoparticle provides the point spread function (PSF) of the endoscope. We find a lateral resolution (PSF full width at half maximum) of 0.83 µm (**Fig. 4d**), while the axial resolution is 5.9 µm (**Fig. 4e**). Both values are in good agreement with the expected NA=0.45 for focusing on the sample. Additionally, we checked that the PSF is not significantly altered when the FoV is increased up to 350 µm (**Fig. S9**). This leads to a remarkable flatness of field as illustrated on the USAF test chart (**Fig. S10**).

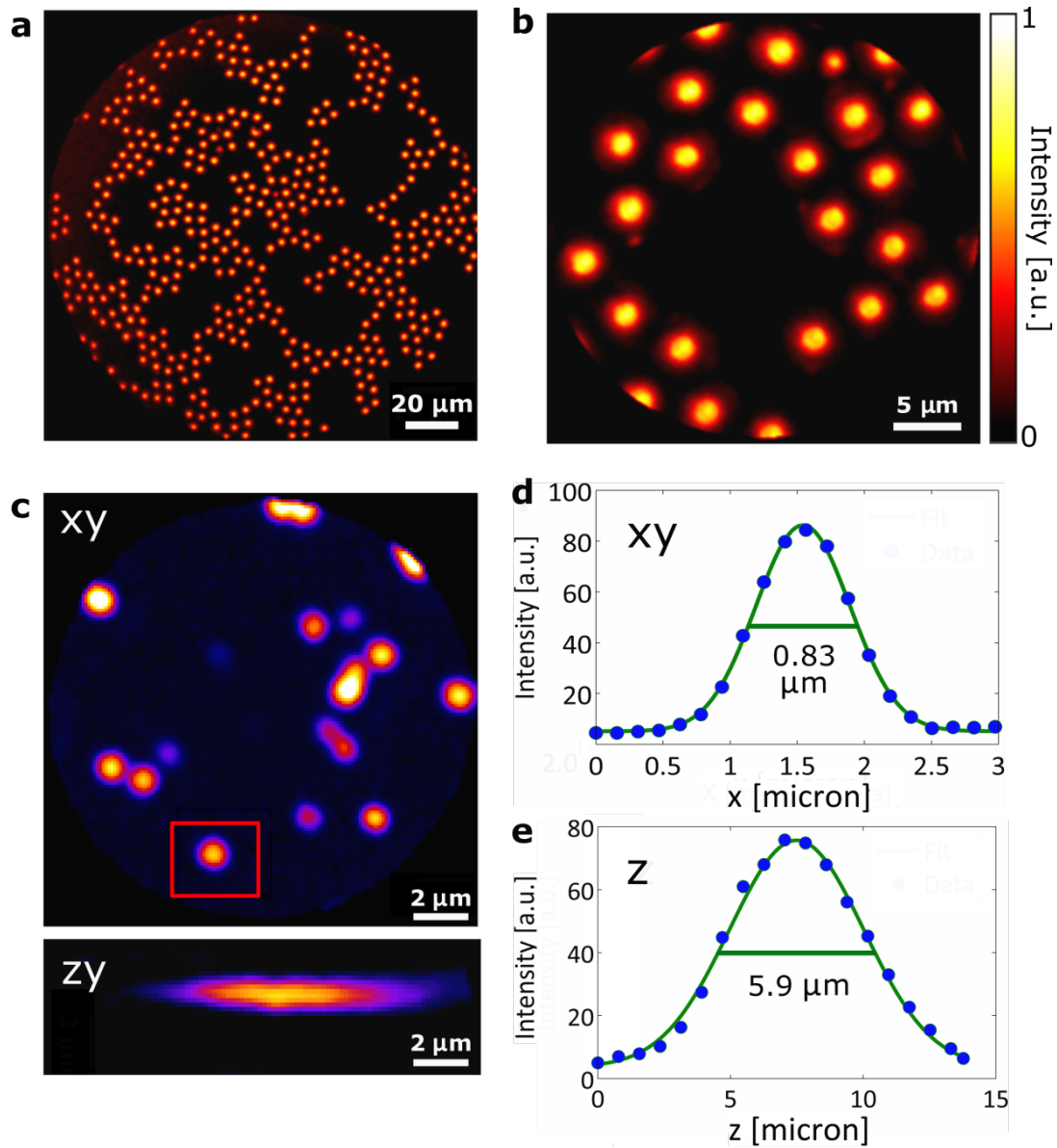

**Figure 4: High-contrast submicron image resolution.** (a, b) CARS images of 5 μm diameter polystyrene beads deposited on a glass coverslip with two different FoVs: 155 μm (a) and 25 μm (b) corresponding respectively to 7.5 V and 1.5 V pk-pk driving voltages. Note the vanishing background between the beads. Powers: 10 mW (pump) and 5 mW (Stokes). Excitation wavelengths: 800 nm (pump) and 1040 nm (Stokes) (c) Lateral (x,y) and axial (x,z) 2-photon fluorescence images (forward detected) of 200 nm diameter fluorescent nanoparticles. Excitation: 800 nm, 10 mW. Detection: 550/100 nm. The nanoparticle highlighted in the red rectangle is used to obtain the image cross-cuts (d,e) from which the lateral and axial PSFs are deduced. The values on the graph indicate the full widths at half maximum (FWHM).

**Endoscopic tissue imaging**

We now move to label-free tissue samples to demonstrate the ability for CARS and SHG multimodal nonlinear endoscopy imaging. **Figure 5a** shows a CARS image from a fresh human colon fatty tissue sample, where the high lipid content provides CARS contrast through the C-H stretch vibrations. Again as in **Fig. 4a, b**, we find an excellent CARS contrast together with a high spatial resolution. We also demonstrate that a single acquisition performed in only 0.8 s leads to a good quality CARS image (**Fig. S11** and **S12**), with only a minor reduction in the signal-to-noise ratio as compared to the average over 5 acquisitions used for **Fig. 5a**.

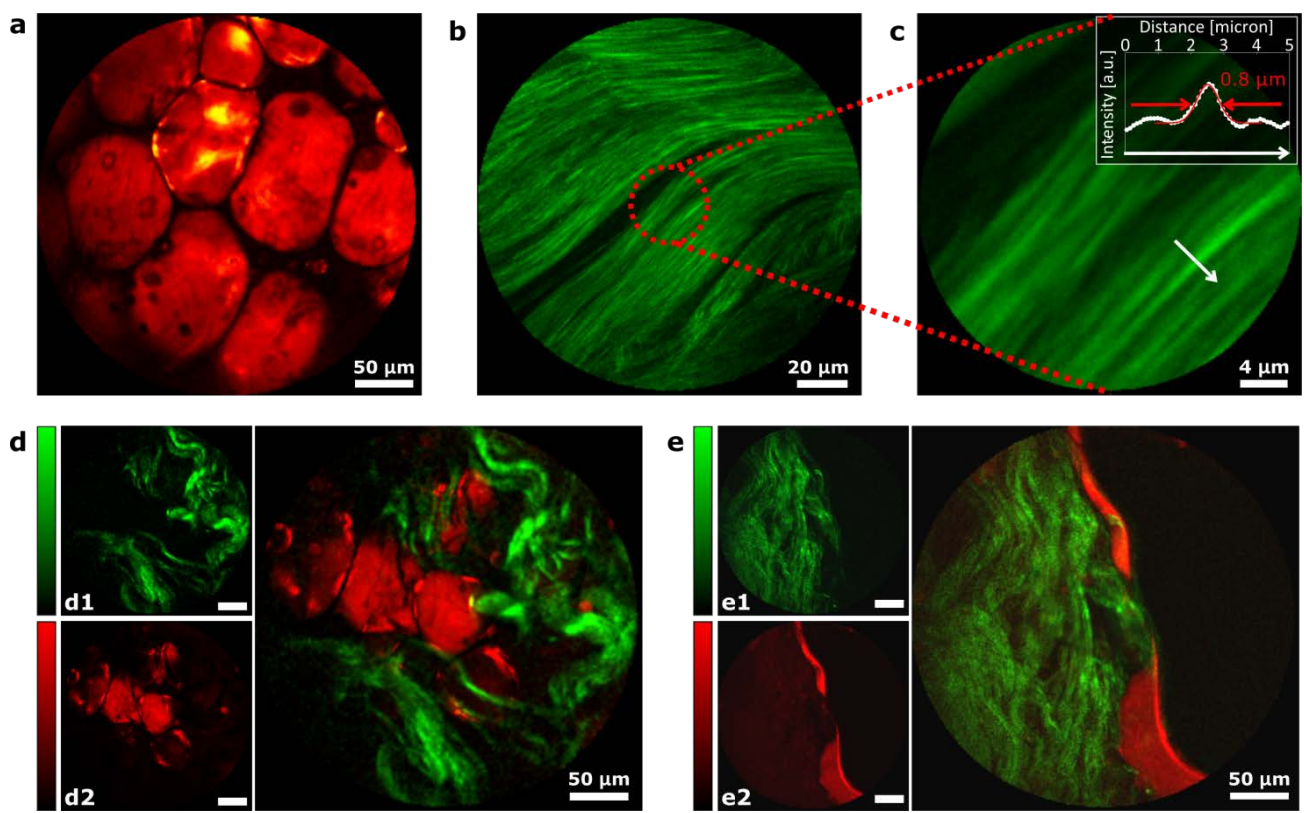

**Figure 5: Multimodal nonlinear endoscopy on biological tissues.** (a) CARS image of fresh fatty tissues from human colon. (b) SHG image of rat tail tendons featuring collagen fibers. (c) close-up view of (b) confirming the 0.8 μm lateral resolution for nonlinear imaging. (d,e) Multimodal SHG (green, d1, e1) and CARS (red, d2, e2) nonlinear images of fresh human colon tissues taken 50 μm below the sample surface. The large images are overlaps of the SHG and CARS images. Laser powers: pump 20 mW, Stokes 10 mW, SHG 60mW (45 mW for images b and c). Excitation wavelengths: 800 nm (pump) and 1040 nm (Stokes). All images were averaged over 5 acquisitions, for a total acquisition time of 6.5 seconds, except b and c that were averaged 10 times (13 seconds).

**Figure 5b** demonstrates the SHG imaging performance on collagen fibers from rat tail tendon. The close-up view of the fibril cross section (**Fig. 5c**) confirms the 0.8 μm lateral resolution of the flexible nonlinear endoscope. Lastly, we merge the CARS and SHG imaging into the multimodal nonlinear images in **Fig. 5d, e** and **Fig. S12.** The samples are fresh tissues from human colon coming directly from surgery without any preparation or staining. The images presented in **Fig. 5 d, e** have been obtained ~50 μm below the sample surface. The CARS images reveal the lipidic content, while the SHG images are sensitive to the presence of collagen. The merged nonlinear images reveal the complex morphological structure of the tissue, highlighting a clear separation between zones enriched in lipids and collagen fibers. We verified that moving the fiber during acquisition doesn't affect image quality such that the distal head can be oriented at will in relation with the tissue sample. Altogether, these results demonstrate the ability of the fiber endoscope to perform multimodal nonlinear imaging directly on fresh tissue that has not been prepared in any way.

**Discussion**

The major challenge faced by coherent Raman endoscopy is to achieve simultaneously a high spatial resolution, a large FoV and a high CARS contrast. Earlier attempts using solid-core fibers have been limited by the strong FWM background occurring inside the solid core[7, 8, 9], with negative impacts on the CARS image contrast quality and/or the bulkiness of the distal endoscope head[7, 11, 13]. Our double clad Kagomé-lattice hollow core fiber solves all these issues by providing negligible FWM background due to its low nonlinearity[22] and very efficient signal back collection. When combined with the microsphere inserted in its distal core the fiber shows a micron size distal focus (**Fig. 3d**) enabling a large FoV operation - thanks to the close to unity magnification of the four elements miniature objective (**Fig. S1**). Furthermore low GVD of the HC fiber enables the direct delivery of sub 200 fs pulses avoiding the complexity associated with dispersion pre-compensation schemes[23]. Finally the Kagomé lattice HC design provides a broad transmission window[24] suitable to address with CARS the high frequency ~3000 $cm^{-1}$ vibrational bonds found in lipids. This is in sharp contrast with earlier work by our group[25] using HC band gap fibers whose narrow transmission window limits the detectable Raman shifts below 1000 $cm^{-1}$.

With an external diameter of 4.2 mm, our endoscope distal head is small enough to be inserted into the user channel of many conventional endoscopes. Further improvements could be made in terms of miniaturization as smaller piezo-tubes might be used to reduce both the rigid length and the external diameter of the probe. Recently, 2 mm outer diameter and 10 frames/s acquisition speed have been demonstrated using a similar resonant piezo scanning technology [26]. Our HC based endoscope is also geared for future SRS imaging developments owing to the fact that parasitic SRS noise in HC Kagomé lattice fiber is $10^4$-$10^6$ lower than in solid core fiber and can be even completely suppressed[27].

**Conclusion**

We have reported a fiber endoscope for high-resolution multimodal nonlinear imaging merging several innovations: hollow-core fiber with a Kagomé lattice, dual fiber clad, microsphere focusing and resonant miniature piezo scanner. This unique combination of key technologies solves most of the issues raised in multi-modal nonlinear endoscopy: ultra-short pulse delivery, efficient signal collection, large field of view, sub-micron resolution, low noise and compactness. High-contrast and high-resolution CARS and SHG nonlinear endoscopy have been demonstrated on unstained tissues using with 0.8 µm resolution over a FoV up to 320 µm. This endoscopy platform is highly versatile and fully compatible with recent coherent Raman laser source developments[28, 29] and with any nonlinear imaging technique using wavelengths within the HC broad transmission window. This includes 2-photon excited fluorescence, SHG, THG, CARS, SRS and also the broad variety of pump-probe schemes[30]. This endoscope platform opens interesting perspectives for intra-operative label free imaging for real-time histopathology diagnosis[5].

**Methods**

Two synchronized 80 MHz femtosecond pulse trains (pump 800 nm – 100 fs, Stokes 1040 nm – 160 fs) are delivered by a fs laser system (Coherent, Discovery). The beam powers were controlled with half-waveplates and polarizing beamsplitters. The temporal delay between the pulses was adjusted by means of a retroreflector mounted on a mechanical translation stage (1 µm steps). The beams were spatially combined with a notch filter optimized for the reflection of the 1040 nm beam (Semrock, NFD01-1040). The beams were injected in the fiber with a 40x microscope objective (NA=0.6, LUCPLFLN 40x, Olympus). The high NA of the lens allowed to collect the entire signal back-collected by the fiber double clad (NA=0.5 at 400 nm). The use of a telescope ($f_1$=400 mm, $f_2$=40 mm) was necessary to reduce the diameter of the excitation beams before injection in the fiber core owing to its low (0.02) numerical aperture. The excitation beams propagated in 1 m of the Kagomé DC fiber, whose distal end was attached to the four-quartered piezo-tube (PI Ceramic, PT230.94). The silica micro-bead (Thermo Scientific, 9000 Series Glass Particle Standards, 30 µm) attached to the fiber core focuses the beams to a micron sized spot. This spot is re-imaged on the sample by means of a 4-lenses (Edmund optics) miniature objective described in the supplementary section (Fig. S1). The nonlinear signals generated by the sample were back-collected by the miniature objective and coupled in the Kagomé DC fiber, to be detected on the endoscope proximal side. A long-pass dichroic beamsplitter (Semrock, FF757-Di01) was used to separate the detection path (<750 nm) from the excitation path (>770 nm). The detection path was composed of the coupling objective lens (NA=0.6, LUCPLFLN 40x, Olympus - f=4.5 mm), a telescope lens (f=40 mm), a 100 mm lens and a 50 mm tube lens. Bandpass filters placed in front of the photomultiplier tube (Hamamatsu, H7421-40) allowed to separate the photons generated by each nonlinear process (2-photon, CARS, SHG).

For calibration purpose only, a microscope objective (Olympus, LUCPLFLN 40x) placed in front of the endoscope distal head allows (i) to measure the spot position during a scan with a position sensitive detector (Thorlabs, PDP90A) and (ii) to image the distal end of the fiber on a CCD camera (Thorlabs, DCC1545).

The resonant fiber-scan and signal detection were controlled by a 1 MHz data acquisition board (National Instruments, NI USB-6151). A custom Labview software synchronized the detection and scan tasks by means of a start trigger and the internal clock of the board (details of the synchronization algorithm can be found in[31]). The images were then reconstructed by means of an open-loop algorithm (described in the supplementary section Fig. S4), that displayed the nonlinear images in real-time in the user-friendly scan-software interface.

The samples used for nonlinear image calibration in Fig. 4 consisted of fluorescent beads (Thermo Fisher Scientific, Carboxylate-Modified Microspheres, 0.2 µm, yellow-green fluorescent (505/515)) and polystyrene beads (Sigma-Aldrich, 79633-5ML-F).

The human colon tissues were sandwiched between two coverslip and imaged with no further staining or preparation. The images presented on Fig. 5 were obtained with the endoscope distal head rigidly mounted on 3D translation stage.

## Acknowledgements

The authors thank F. Poizat, F. Caillol and M. Giovannini for giving access to the tissue samples. A.K. acknowledge assistance from K. Delplace, C. Fourcade-Dutin, R. Habert and D. Labat. This work was supported by EU-ITN-607842-2013-FINON, FR-"Investissement d'Avenir"-11-IDEX-0001-02, 11-INSB-0006, 11-EQPX-0017, 11-LABX-0007,FR-ANR-14-CE17-0004-01, FR-INSERM-PC201508, EU Regional Development Fund (ERDF); Centre National de la Recherche Scientifique (IRCICA USR 3380). X. C. is grateful to the Chinese Science Council (China) for funding support.

## Author contributions

A.L., V.M., S.S. and X.C. performed all experiments. H.R, E.R.A., A.K conceived and designed all experiments. A.L., V.M, S.S., X.C., H.R analyzed the data. A.L. developed the imaging software. J.W. conceptualized the microsphere focusing scheme. A.K and N.J. conceptualized and made the fiber. F.L. and M.F. designed and perform the distal endoscope head integration. H.R, A.L., E.R.A, J.W. wrote the manuscript.

## Competing financial interests

The authors declare no competing financial interests.

# Supplementary Information

# High-resolution multimodal flexible coherent Raman endoscope

Alberto Lombardini[1], Vasyl Mytskaniuk[1], Siddharth Sivankutty[1], Esben Ravn Andresen[1,2], Xueqin Chen[1], Jérôme Wenger[1], Marc Fabert[4], Nicolas Joly[3], Frédéric Louradour[4], Alexandre Kudlinski[2], and Hervé Rigneault[1*]

[1] Aix Marseille Univ, CNRS, Centrale Marseille, Institut Fresnel, Marseille, France

[2] Université Lille, CNRS, UMR 8523, Laboratoire de Physique des Lasers Atomes et Molécules, F-59000 Lille, France

[3] Max Planck Institute for the Science of Light, and Department of Physics, University of Erlangen Nuremberg, 91058 Erlangen, Germany

[4] Université de Limoges, CNRS, XLIM, UMR 7252, Université de Limoges, F-87060 Limoges, France

[*]herve.rigneault@fresnel.fr

**Content:**

- **Figure S1**: Miniature objective lens
- **Figure S2**: Resonant four-quartered piezo scanner
- **Figure S3**: Resonant piezo scanner calibration and image acquisition
- **Figure S4**: Image reconstruction
- **Figure S5**: HC fibers deliver fs pulses without time and spectral distortions
- **Figure S6**: HC fiber shows negligible four wave mixing (FWM) background for co-propagating ~100fs pump and Stokes pulses
- **Figure S7**: HC fiber enables background-free CARS detection
- **Figure S8**: The HC double clad collects the generated nonlinear signals in the case of scattering samples
- **Figure S9**: PSF off-axis aberrations are negligible for a FoV up to 350 µm
- **Figure S10**: Probe transmission imaging through USAF-1951 resolution chart.
- **Figure S11**: Comparison between CARS averaged images and single acquisition
- **Figure S12**: CARS and SHG multimodal imaging with the flexible nonlinear endoscope

## 1. Miniature objective lens

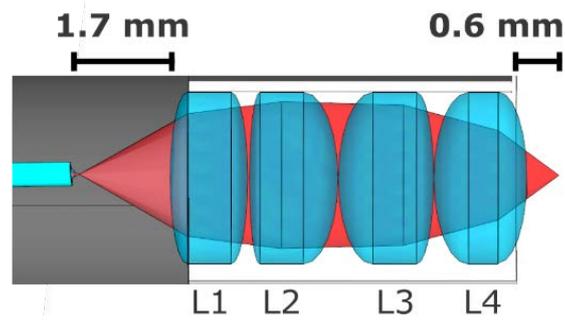

**Figure S1:** Detailed view of the endoscope distal head featuring the miniature objective lens made of four 2 mm diameter achromatic doublets providing a 0.63 magnification. The focal lengths for the lenses are L1: f=9 mm, L2: f=6 mm, L3: f=4 mm, L4: f=3 mm. Each lens has a thickness of 2 mm. Reference of the Edmund Optics (EO) microlenses: L1 - EO #83-338, L2 – EO #65-569, L3 – EO #65-568, L4 – EO #65-567.

## 2. Resonant four-quartered piezo scanner

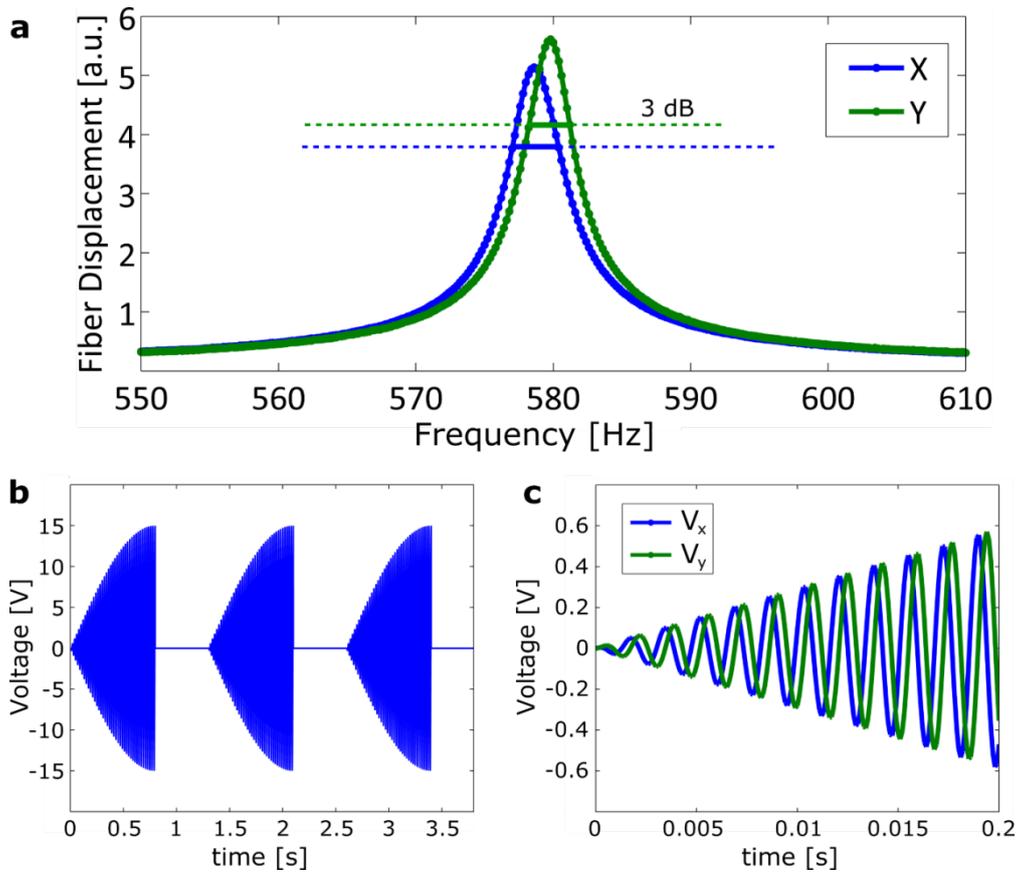

**Figure S2:** (a) Resonant frequency of the piezo X and Y axis with a 23 mm fiber free standing length. (b) Piezo driving voltage as a function of time, the image acquisition is performed during the expanding pattern and takes 0.8 s, which is followed by a 0.5 s rest period; (c) voltages applied along the X ($V_x$) and Y ($V_y$) piezo axis during the first 0.2 s of the expanding pattern showing. The $\pi/2$ phase shift between $V_x$ and $V_y$ leads to the spiral scan.

## 3. Resonant piezo scanner calibration and image acquisition

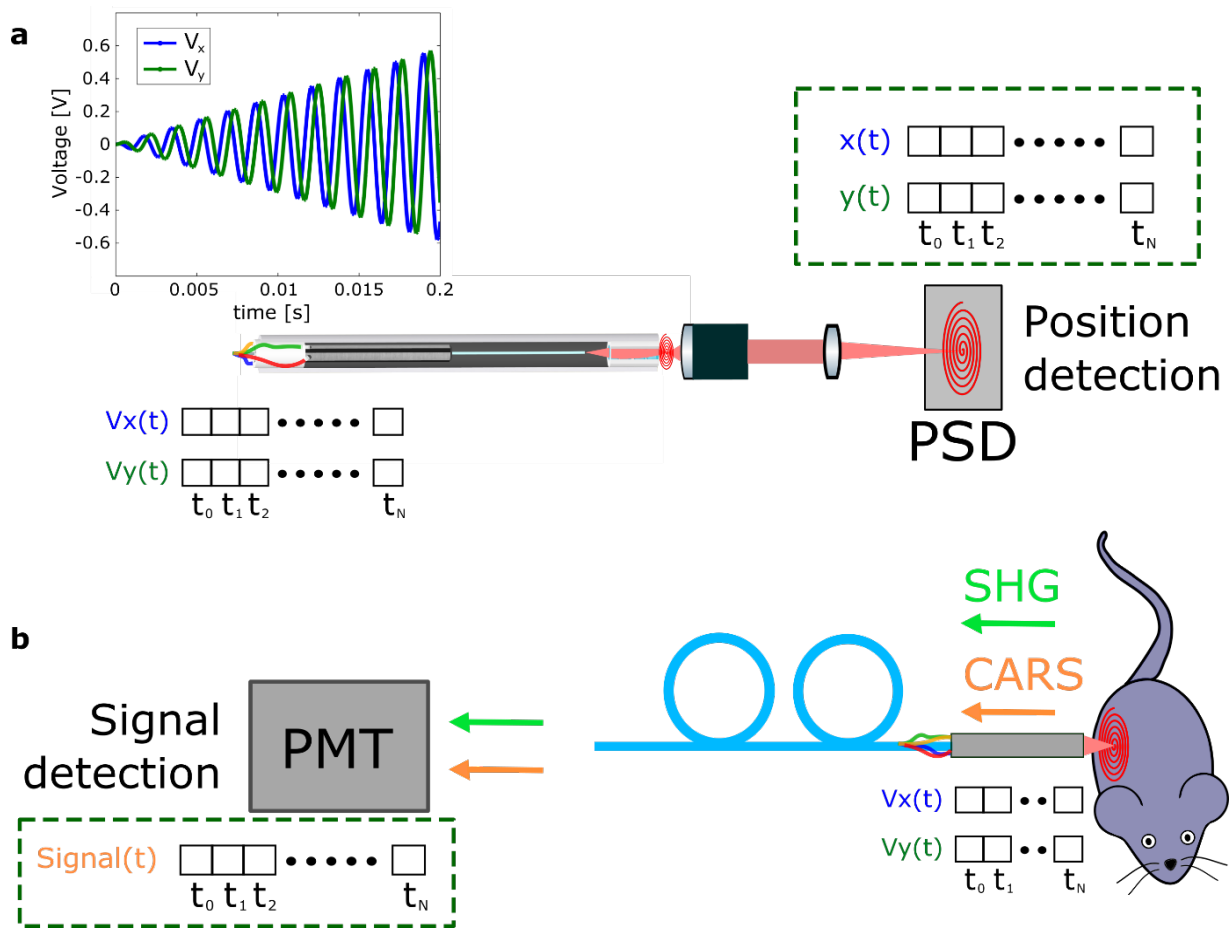

**Figure S3:** Imaging with the fiber scanner requires the measurement of a signal and the knowledge of the instantaneous spot position. The latter is obtained in a first calibration step (a), where the beam position is tracked during a scan with a position sensitive detector (PSD - PDP90A, Thorlabs). When imaging is performed (b), the same waveforms $V_x(t)$ and $V_y(t)$ are applied to the piezo-tube and the nonlinear signals (CARS, SHG, 2-photon) are back-collected by the endoscope distal probe, transmitted by the fiber to the proximal photomultiplier tube (PMT - H7421-40, Hamamatsu) that records the signal $S(t)$.

The positions and signal vectors are then used by the software to reconstruct the images, as shown in **Fig. S4**. A daily calibration with the PSD is used to compensate for small, probably thermal-induced drifts in the scan pattern.

## 4. Image reconstruction

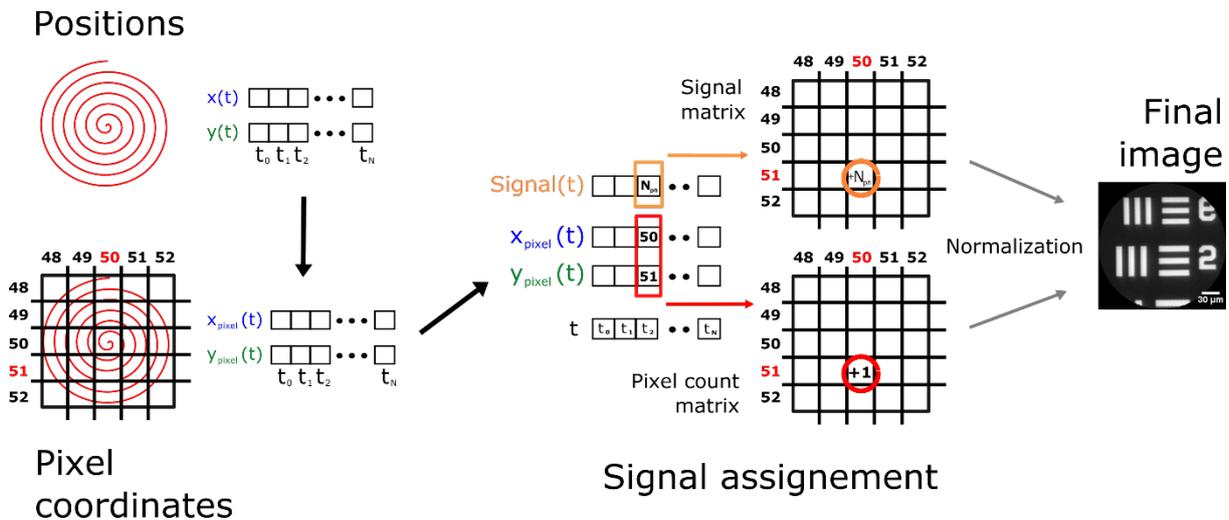

**Figure S4:** The image reconstruction step is performed right after the signal acquisition from a scan. First of all, the spot positions measured in the calibration step (x(t) and y(t), see **Fig. S3**) are mapped into discrete pixel coordinates ($x_{pixel}(t)$ and $y_{pixel}(t)$). These coordinates correspond to the elements of a square matrix whose size is equal to the desired image size (arbitrarily defined by the user). The image is then built from the simultaneous readout of the signal vector and the pixel coordinate vectors. The intensity of the signal detected at each specific time during the scan is assigned to the pixel in the image that corresponds to the spot position at that same time. Since the sampling of the FoV over time is not uniform, a matrix (pixel count) has to be built that contains the number of occurrences that each pixel is sampled. This matrix is then used to normalize the one that contains the signal information. This normalization results in the desired image. Fiber driving and signal acquisition are synchronized and controlled with a custom designed LabView software.

## 5. HC fibers deliver fs pulses without time and spectral distortions

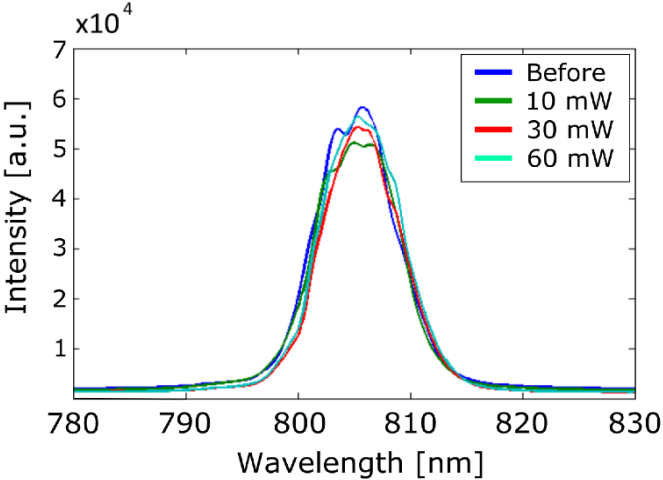

**Figure S5:** Pulse spectrum measurements before and after 1 m of propagation in the HC fiber for different input powers. The weak air filled core nonlinearity does not affect the fs pulse spectrum for the considered power levels.

## 6. HC fiber shows negligible four wave mixing (FWM) background for co-propagating ~100fs pump and Stokes pulses

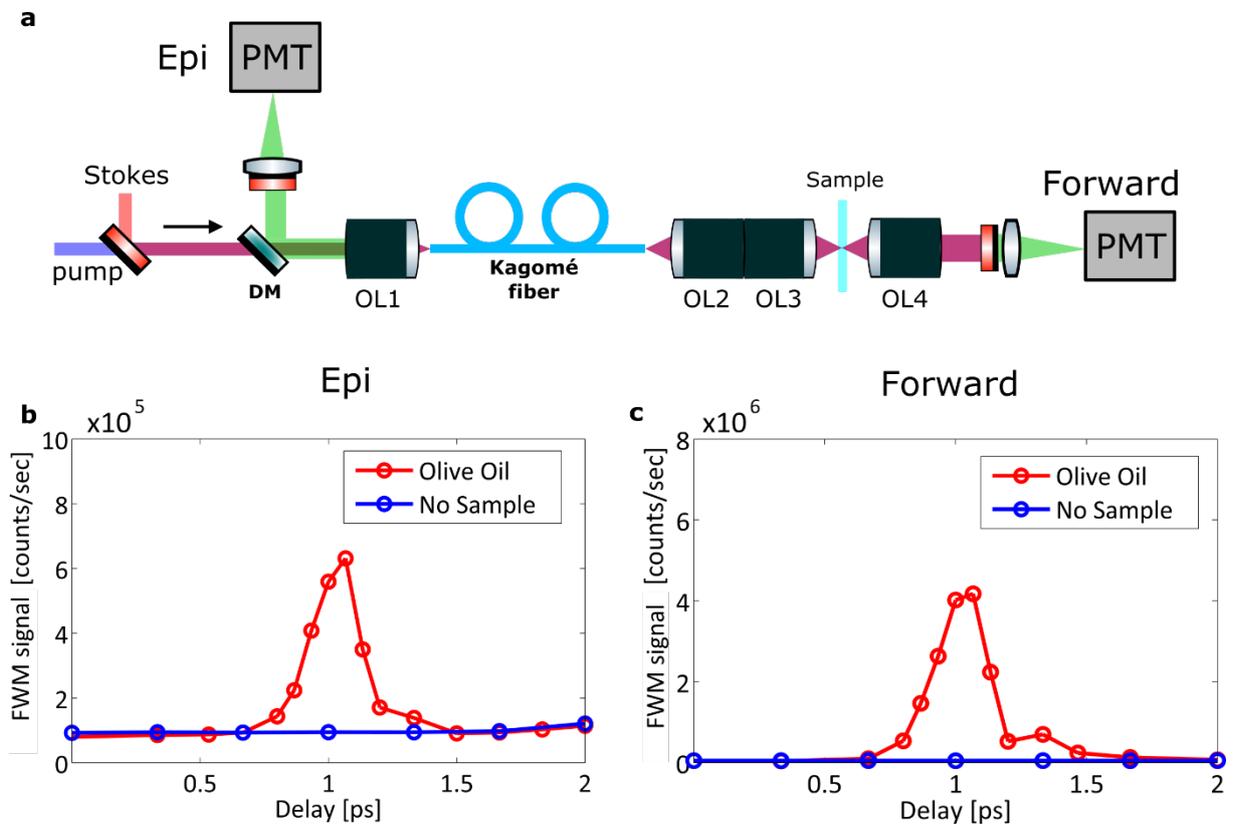

**Figure S6:** (a) setup of the experiment, two back-to-back 40x, NA=0.6 objective lenses (OL2 and OL3) are used to focus the pump (780 nm – 100 fs) and Stokes (1031 nm – 100 fs) beams at the sample plane on a cuvette with (or without) olive oil; (b) epi and (c) forward detected CARS signal with and without the olive oil sample as a function of the delay between the pump and Stokes pulses. For an optimized pump and Stokes beam temporal overlap at the sample plane (delay 1 ps) the generated HC fiber FWM signal is negligible. DM: dichroic mirror, OL: objective lens, PMT: photomultiplier tube. OL1 and OL4 = Olympus LUCPLFLN 20x, NA=0.45. OL2 and OL3 = Olympus LUCPLFLN 40x, NA=0.6. All measurements are performed with a FemtoFiber dichro bioMP (Toptica, Munich), power at the sample plane are $P_{pump}$=12 mW (780 nm) and $P_{Stokes}$=8 mW (1031 nm).

## 7. HC fiber enables background-free CARS detection

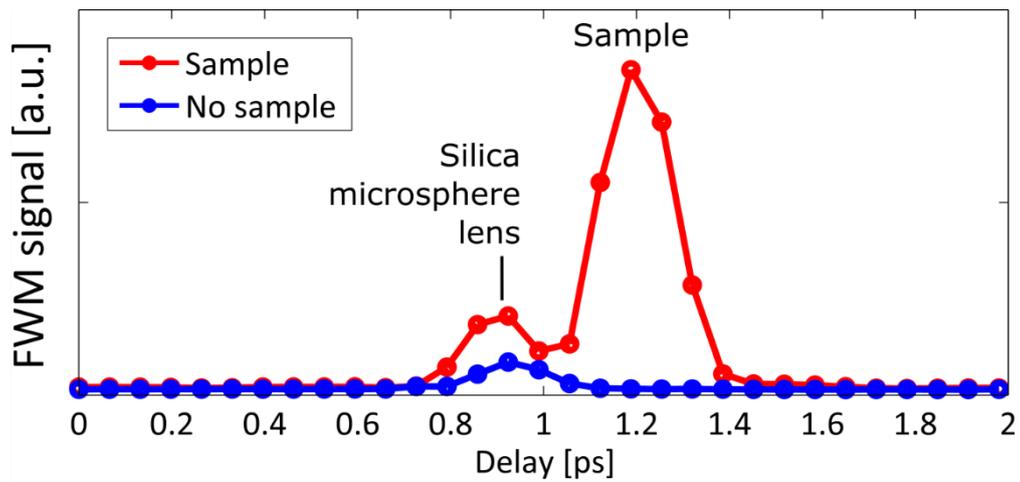

**Figure S7**: CARS signal detected by the flexible endoscope probe when scanning the delay between the pump and the Stokes pulses. In the absence of the sample (blue curve), a weak FWM background is found at 0.9 ps delay between the pump and Stokes pulses. We attribute this background to the silica microsphere lens inserted into the fiber core. In the presence of the olive oil sample (red curve), this background is still present but is clearly separated from the strong CARS signal from the lipids arising at 1.2 ps delay. This delay difference is attributed to the inherent dispersion of the miniature objective. Importantly, the background without the sample is vanishing for the 1.2 ps delay leading to optimum pump and Stokes pulses overlap in the sample plane.

## 8. The HC double clad collects the generated nonlinear signals in the case of scattering samples

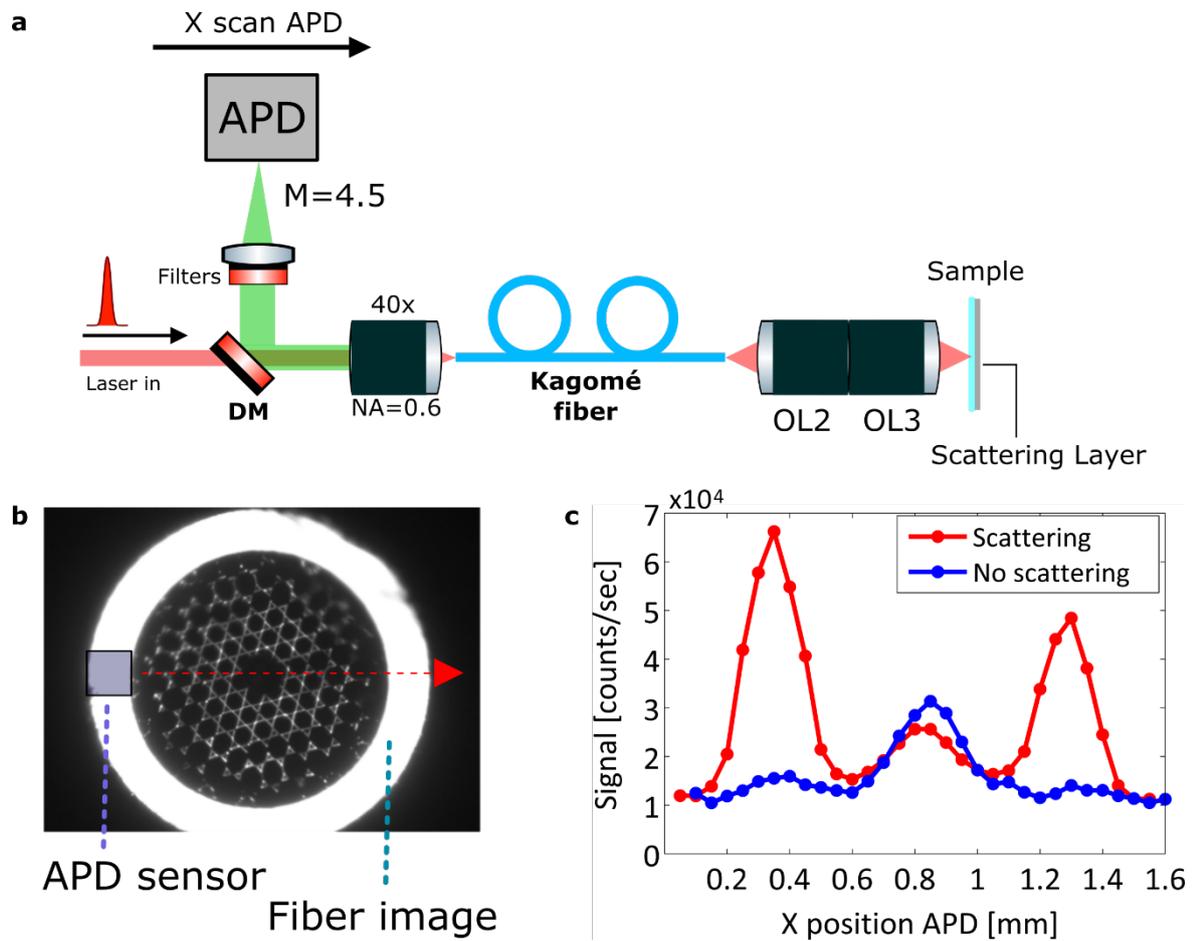

**Figure S8:** (a) setup of the experiment. The sample is a fluorescein (FITC) cuvette solution excited at 900 nm. The 2-photon fluorescence signal collected by the HC fiber is filtered in the 500 – 550 nm band and analyzed spatially by scanning an APD with a small 150 µm active surface (SPCM, Perkin Elmer). For this, the output facet of the HC-DC fiber is magnified (x4.5) and imaged onto the APD; (b) output facet of the HC-DC fiber showing the de-magnified image of the APD active surface and the performed linear scan (red dash line). (c) Collected 2-photon signal across the HC-DC output facet when the back surface of the FITC cuvette is made clear (blue) and when it is covered with a thin layer of $TiO_2$ scattering nanoparticles (200 nm diameter – layer thickness 500 µm) (red). Most of the back emitted 2-photon signal is collected by the DC in the case of a scattering sample whereas it is collected by the air core in the case of a non-scattering sample. All measurements are performed with a Chameleon (Coherent) Ti:Saph laser.

## 9. PSF off-axis aberrations are negligible for a FoV up to 350 µm

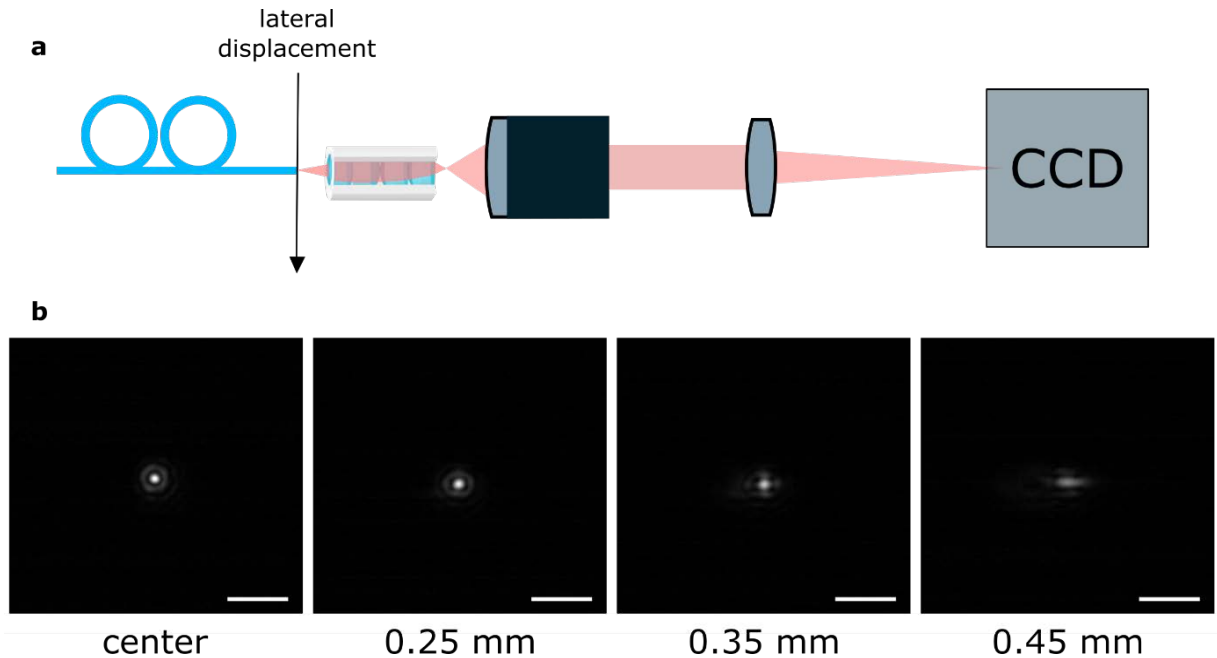

**Figure S9:** (a) PSF evolution when the HC-DC fiber distal facet is laterally displaced from the optical axis of the miniature objective (the magnification between the HC-DC distal facet and the CCD camera is 33); (b) evolution of the PSF with lateral displacement – scale bar is 10 µm. The beam profile is not significantly affected, up to a lateral displacement of 250 µm (FoV 500 µm). However, when the lateral displacement reaches 350 µm the beam profile becomes slightly asymmetric, up to an elongated profile observed at 450 µm. We conclude that the probe has a uniform PSF across a FoV of 320 µm (corresponding to 500 µm piezo fiber scan - with a de-magnification of 0.63). All measurements are performed with a Chameleon (Coherent) Ti:Saph laser operating at 800 nm.

## 10. Probe transmission imaging through USAF-1951 resolution chart.

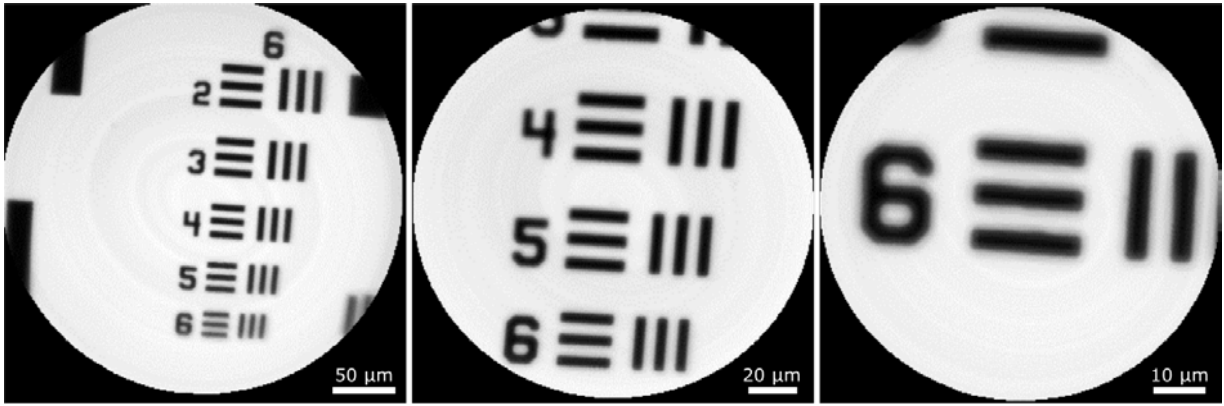

**Figure S10:** Probe transmission imaging through a USAF-1951 resolution chart with driving voltages ±15 V (30 V pk-pk) (a), ±7.5 V (b) and ±3.75 V (c). The experiments were performed by recording the transmitted signal with a photodiode (DET10A, Thorlabs) as the laser spot was scanned across the transmission chart.

## 11. Comparison between CARS averaged images and single acquisition

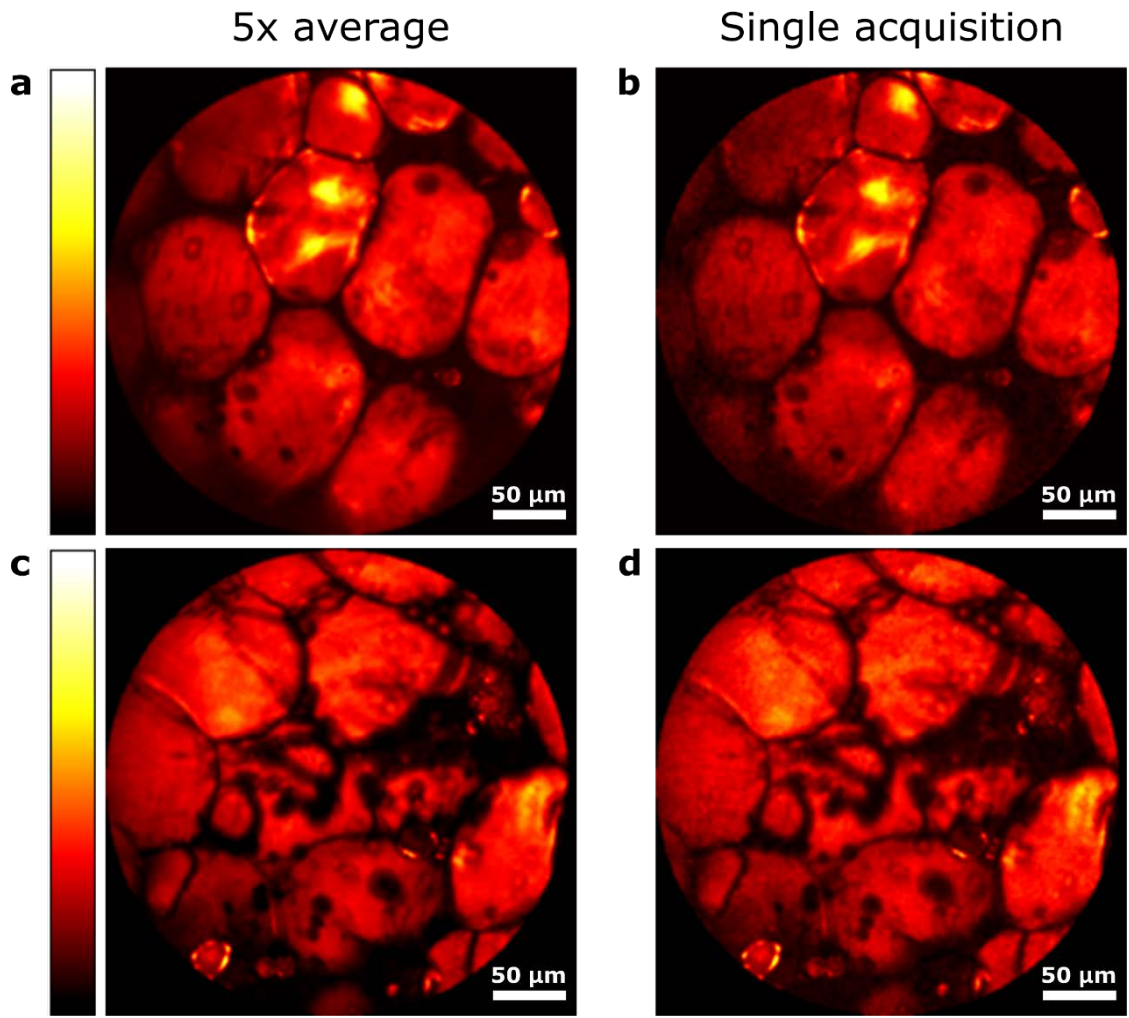

**Figure S11:** Comparison of CARS images from a human colon tissue with 5x averaging / 6.5 s acquisition time (a, c) and a single 0.8 s acquisition (b, d). Power on the sample: $P_{pump}$=20 mW, $P_{Stokes}$=10 mW. All measurements are performed with a Discovery (Coherent) operating at 800 nm and 1040 nm.

## 12. CARS and SHG multimodal imaging with the flexible nonlinear endoscope

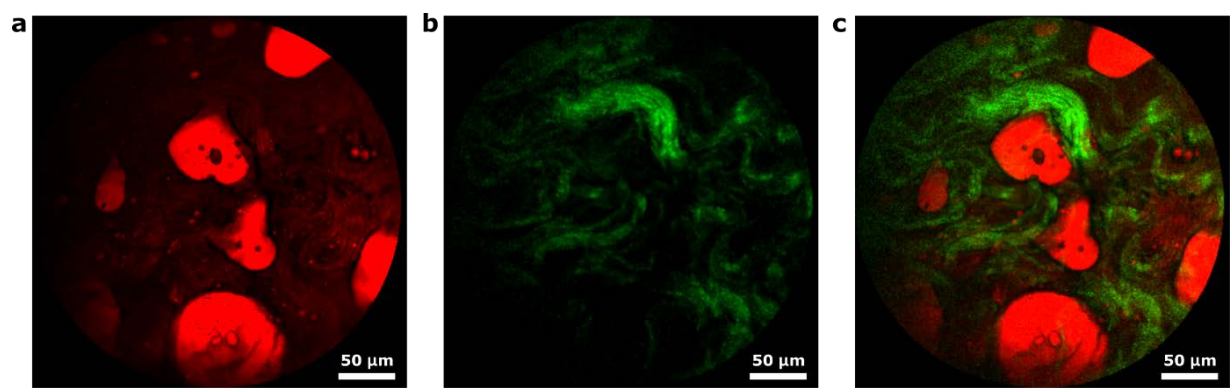

**Figure S12:** Multimodal nonlinear image in fresh human colon. (a) CARS image, (b) SHG image, (c) overlay - CARS powers: pump 20 mW, Stokes 10 mW; SHG powers: pump 60mW; all images averaged over 5 acquisitions (6.5 s total acquisition time).